\documentclass[aps,prx,reprint,superscriptaddress,intlimits]{revtex4-2}
\usepackage{bm,latexsym,mathrsfs,enumerate,amsmath,amssymb}

\usepackage{graphicx}
\usepackage[breaklinks=true,urlcolor = blue,colorlinks = true,citecolor = blue,linkcolor = blue]{hyperref}

\begin{document}

\title{Three-dimensional Topological Superstructure of Magnetic Hopfions Threaded by Meron Strings in Easy-plane Magnets}

\author{Shoya Kasai}
  \email{Contact author: kasai@aion.t.u-tokyo.ac.jp}
  \affiliation{Department of Applied Physics, The University of Tokyo, Tokyo 113-8656, Japan}
\author{Kotaro Shimizu}
  \affiliation{RIKEN Center for Emergent Matter Science, Saitama 351-0198, Japan}
\author{Shun Okumura}
  \affiliation{Department of Applied Physics, The University of Tokyo, Tokyo 113-8656, Japan}
\author{Yasuyuki Kato}
  \affiliation{Department of Applied Physics, University of Fukui, Fukui 910-8507, Japan}
\author{Yukitoshi Motome}
  \email{Contact author: motome@ap.t.u-tokyo.ac.jp}
  \affiliation{Department of Applied Physics, The University of Tokyo, Tokyo 113-8656, Japan}

\begin{abstract}
Topological spin textures exhibit a hierarchical nature. For instance, magnetic skyrmions, which possess a particle-like nature, can aggregate to form superstructures such as skyrmion strings and skyrmion lattices. Magnetic hopfions are also regarded as superstructures constructed from closed loops of twisted skyrmion strings, which behave as another independent particles. However, it remains elusive whether such magnetic hopfions can also aggregate to form higher-level superstructures. Here, we report a stable superstructure with three-dimensional periodic arrangement of magnetic hopfions in a frustrated spin model with easy-plane anisotropy. By comprehensively examining effective interactions between two hopfions, we construct the hopfion superstructure by a staggered arrangement of one-dimensional hopfion chains with Hopf number $H=+1$ and $H=-1$ running perpendicular to the easy plane. Each hopfion chain is threaded by a magnetic meron string, resulting in a nontrivial topological texture with skyrmion number $N_{\rm sk}=2$ per magnetic unit cell on any two-dimensional cut parallel to the easy plane. We show that the hopfion superstructure remains robust as a metastable state across a range of the hopfion density. Furthermore, we demonstrate that superstructures with higher Hopf number can also be stabilized. Our findings extend the existing hierarchy of topological magnets and pave the way for exploring new quantum phenomena and spin dynamics.
\end{abstract}

\maketitle
% \tableofcontents

%%%%%%%%%%%%%%%%%%%%%%%%%%%%%%%%%%%%%%%%%%%%%%%%%%%%%%%%%%%%%%%%%%%%%%%%%%%%%%%%%%%%%%%%%%%%%%%%%%%%%%%%%%%%%%%%%%%%%%%%%%%%%%%%%%%%
\section{Introduction}
\label{sec:intro}
Topology is a branch of mathematics that classifies geometric properties that remain invariant to continuous deformations. For instance, consider mapping a geometric structure onto an $n$-dimensional spherical surface $S^n$ by means of a continuous map. The mapped image may cover $S^n$ an integer number of times. This integer value, called the winding number, distinguishes the topological properties of the original geometric structure. Structures characterized by different winding numbers can never be mutually transformed by continuous deformations. In this sense, such a topological invariant endows the structure with substantial stability, known as topological protection.

The concept of topology has also been applied to condensed matter physics, and led to many breakthroughs to date. It was initiated by the quantum Hall effect, i.e., quantization of the Hall conductivity in two-dimensional (2D) electron systems under an external magnetic field~\cite{Klitzing1980}. This surprising phenomenon originates from a nonzero topological invariant called the Chern number, which characterizes the topology of the electronic band structure in momentum space~\cite{Laughlin1981,Thouless1982}. This concept was extended to a spontaneous quantum anomalous Hall effect that does not require a magnetic field~\cite{Haldane1988}, leading to the discovery of topological insulators~\cite{Kane2005,Bernevig2006,Konig2007}, which have been the subject of intense experimental and theoretical research in recent years.

%%%%%%%%%%%%%%%%%%%%%%%%%%%%%%%%%%%%%%%%%%%%%%%%%%%%%%%%%%%%%%%%%%%%%%%%%%%%%%%%%%%%%%%%%%%%%%%%%%%%%%%%%%%%%%%%%%%%%%%%%%%%%%%%%%%%
\begin{figure*}[tb]
    \centering
    \includegraphics[width=\hsize]{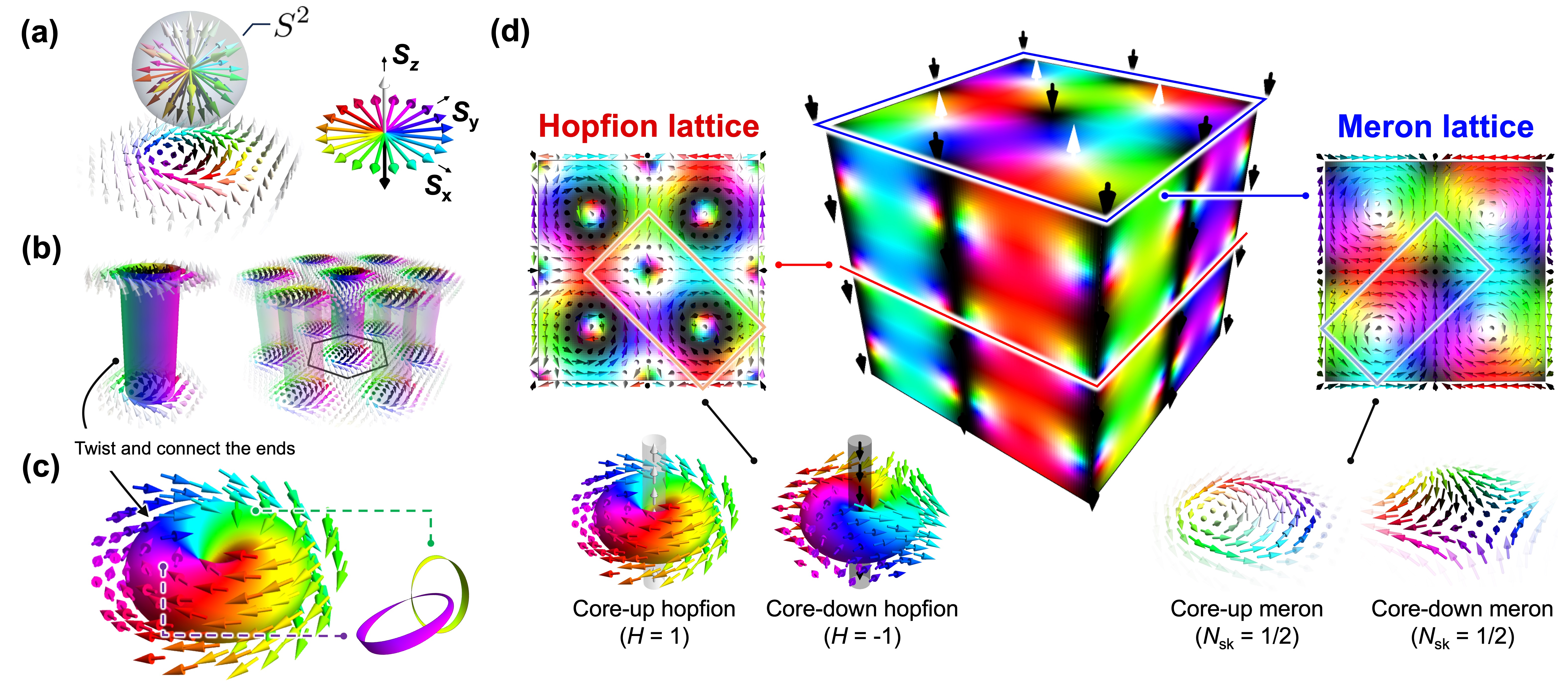}
    \caption{Schematic illustrations of hierarchy of topological spin textures. (a) Magnetic skyrmion and the stereographic projection of its spin configuration onto a 2D spherical surface $S^2$. The arrows represent spins whose $S_x$ and $S_y$ ($S_z$) components are indicated by color (gray scale), as shown in the inset. This color code is used throughout this paper. (b) Skyrmion string (left) and skyrmion lattice (right). The black hexagon in the right panel denotes the 2D magnetic unit cell containing a single skyrmion. (c) Magnetic hopfion and two isospin lines linked with each other. (d) 3D hopfion superstructure with a periodic arrangement of vertical chains of hopfions. It is composed of staggered arrangement of hopfion chains with Hopf number $H=+1$ and $-1$; the white and black arrows indicate cores of the hopfion chains with up and down spins, respectively. The spin textures on two horizontal cuts are shown in the left and right insets. The left inset is the cut through the centers of hopfions, where the black and white rings represent the cuts of hopfions with Hopf number $H=+1$ and $-1$ as shown below. The right one is the cut between stacked hopfions, regarded as a magnetic meron lattice of core-up and core-down merons with skrmion numbers $N_{\mathrm{sk}}=1/2$ as shown below.}
    \label{fig:hierarchy}
\end{figure*}
%%%%%%%%%%%%%%%%%%%%%%%%%%%%%%%%%%%%%%%%%%%%%%%%%%%%%%%%%%%%%%%%%%%%%%%%%%%%%%%%%%%%%%%%%%%%%%%%%%%%%%%%%%%%%%%%%%%%%%%%%%%%%%%%%%%%

Topology in condensed matter physics is also found in real space. A typical example is topological spin textures. The magnetic skyrmion shown in Fig.~\ref{fig:hierarchy}(a) is the prototype. The swirling noncoplanar spin configuration can wrap a 2D spherical surface $S^2$ an integer number of times through stereographic projection. Thereby the magnetic skyrmion is a topological spin structure characterized by a topological invariant called the skyrmion number $N_{\rm sk}$. The example in Fig.~\ref{fig:hierarchy}(a) shows the skyrmion with $N_{\rm sk} = -1$. 

Such topological spin textures exhibit an intriguing hierarchical structure. For instance, the 2D magnetic skyrmions can assemble in the perpendicular direction to form a one-dimensional (1D) string-like superstructure called the skyrmion string, as shown in the left panel of Fig.~\ref{fig:hierarchy}(b)~\cite{Milde2013}. Furthermore, the skyrmion strings can also aggregate to form a 2D periodic superstructure called the skyrmion lattice, as shown in the right panel of Fig.~\ref{fig:hierarchy}(b). These exotic structures are indeed observed in various magnets, such as noncentrosymmetric systems with breaking inversion symmetry~\cite{Bogdanov1989,Roßler2006,Muhlbauer2009,Yu2010,Yu2011,Seki2012,Kézsmárki2015,Kurumaji2017,Tokura2021}, frustrated spin systems~\cite{Okubo2012,Leonov2015}, magnetic metals with long-range and multiple-spin interactions originating from itinerant electrons~\cite{Akagi2010,Akagi2012,Ozawa2016,Ozawa2017,Hayami2017,Okumura2020,Okumura2022,Kurumaji2019,Hirschberger2019,Khanh2020,Takagi2022,Yoshimochi2024}, and magnets with short-range multiple-spin interactions~\cite{Kurz2001,Heinze2011,Grytsiuk2020}.

At each level of the hierarchy, the topologically-protected spin textures exhibit unique emergent electromagnetic properties, and thus hold the promise of applications. The skyrmions generate emergent magnetic fields from the noncoplanar spin texture via the Berry phase mechanism and can be moved by an ultralow electric current, allowing them to be utilized as nonvolatile information carriers on a racetrack~\cite{Fert2013} and a logic device~\cite{Zhang2015}. When the skyrmions form a lattice, the emergent magnetic fields can coherently influence macroscopic properties, such as the topological Hall effect~\cite{Neubauer2009,Nagaosa2013}, the Nernst effect~\cite{Mizuta2016,Hirschberger2020}, and the magneto-optical Kerr effect~\cite{Kato2023}. The skyrmion strings also exhibit unique phenomena, e.g., nonreciprocal nonlinear topological Hall responses induced by string deformations~\cite{Yokouchi2018}, and can work as a waveguide for propagating spin waves~\cite{Seki2020,Xiangjun2020}.

The magnetic hopfion is yet another superstructure composed of a closed loop of twisted magnetic skyrmion string, as shown in Fig.~\ref{fig:hierarchy}(c)~\cite{Sutcliffe2017}. It is characterized by a topological invariant called the Hopf number $H$, derived from the knot theory in mathematics, as follows. Consider the ``isospin line", also referred to as the preimage, on which all spins point in the same direction. In the magnetic hopfion, all the isospin lines form a 1D closed ring, as shown in Fig.~\ref{fig:hierarchy}(c). The Hopf number is defined by how many times any pair of independent isospin lines link with each other. The example in Fig.~\ref{fig:hierarchy}(c) shows the hopfion with $H=1$. This type of intriguing topological structure has been found in high-energy physics~\cite{Faddeev1997}, liquid crystals~\cite{Ackerman2017_nmat,Ackerman2017_PRX,Tai2019}, bosonic systems~\cite{Kawaguchi2008,Hall2016}, ferroelectrics~\cite{Lukyanchuk2020}, and photonic systems~\cite{Sugic2021,Shen2023}. Lately, hopfions in magnets have also garnered attention, with experimental and theoretical studies focusing on their observation~\cite{Kent2021,Yu2023,Zheng2023} and the development of microscopic models realizing them~\cite{Sutcliffe2018,Liu2018,Tai2018,Li2022,Voinescu2020,Bogolubsky1988,Rybakov2022}. It has also been discovered that magnetic hopfions exhibit unique emergent magnetic properties, such as the absence of Hall motion due to cancellation of the total emergent magnetic field~\cite{Wang2019,Liu2020,Liu2022,Saji2023}. This contrasts with the skyrmion Hall effect that hinders the application of magnetic skyrmions on a racetrack~\cite{Jiang2017}

The magnetic hopfions can behave as independent particles, similar to the magnetic skyrmions. Given the hierarchical nature of topological spin textures, it is natural to ask whether they can also aggregate to form superstructures. Indeed, a stable 2D hopfion superstructure was reported in a confined geometry of thin films~\cite{Tai2018}, where the perpendicular magnetic anisotropy near the surfaces of the system plays a crucial role in stabilizing magnetic hopfions~\cite{Sutcliffe2018,Liu2018,Tai2018,Li2022}. However, for the bulk systems without such surface effects, the feasibility of three-dimensional (3D) superstructures of magnetic hopfions has not been fully explored thus far, except for oligomeric assemblies in helical and conical backgrounds~\cite{Voinescu2020}. While the stability of a single hopfion has been studied in bulk frustrated spin systems~\cite{Bogolubsky1988,Rybakov2022,Sallermann2023,Lobanov2023}, how the hopfions aggregate remains elusive. Given that 3D superstructures have been proposed for other physical systems such as liquid crystals~\cite{Tai2019} and bosonic systems~\cite{Li2016}, this is due to the lack of fundamental knowledge about the hopfion-hopfion interactions through microscopic magnetic interactions between spins.

In this paper, we construct a stable 3D superstructure with periodic arrangement of magnetic hopfions in a frustrated spin system and propose experimental methods for detecting it. This is, to our knowledge, the first example of 3D hopfion crystalline superstructures in a bulk magnet. To achieve this, we begin with a comprehensive study of effective interactions between two hopfions under various relative arrangements, performing large-scale energy optimization of spin configurations based on the gradient descent method supported by automatic differentiation. Based on this knowledge, we construct the 3D superstructure step by step. First, we examine the stability of a 1D chain of stacked hopfions and find that the easy-plane magnetic anisotropy plays an important role in the stabilization. Then, we construct the 1D hopfion chains in various forms and discover that a staggered arrangement of hopfion chains with Hopf number $H=+1$ and $H=-1$ in a square-lattice fashion remains stable, resulting in the 3D hopfion superstructure.

We show that in the obtained superstructure, each hopfion chain is threaded by a string of half skyrmions called magnetic merons. Consequently, the 2D cut perpendicular to the hopfion chains is topologically nontrivial, with skyrmion number $N_{\rm sk}=2$ per magnetic unit cell. This peculiar spin configuration generates a net emergent magnetic field, leading to unconventional transport such as the topological Hall effect when coupled with electrons. This result is rather surprising, given the cancellation of the emergent magnetic field for an isolated magnetic hopfion. In addition, we show that the superstructure exhibits higher harmonics in the spin structure factor, characteristic of superpositions of multiple spin density waves. This feature would be useful for identifing the superstructure, e.g., in neutron scattering experiments. 

We carefully examine the stability of the 3D hopfion superstructure while varying the density of hopfions. We show that the superstructure remains robust across a range of the hopfion density, while it is always a metastable state with a slightly higher energy per spin, of the order of $10^{-4}$ of the dominant magnetic interaction strength, compared to the ferromagnetic ground state. Our energy optimization suggests that the superstructure can be self-organized when hopfions are generated with proper density and arrangement in the system. Furthermore, we find that another superstructure with higher Hopf number $H=\pm 2$ can be formed in a certain range of hopfion density. This suggests the possibility of a further variety of hopfion superstructures. 

The rest of the paper is organized as follows. In Sec.~\ref{sec:model}, we introduce the model and the optimization method for spin configurations. In Sec.~\ref{subsec:int}, we present the simulation results for the effective interactions between hopfions in the absence of the easy-plane anisotropy. In Sec.~\ref{subsec:anisotropy}, we show that a 1D hopfion chain is stabilized by introducing the easy-plane anisotropy. In Sec.~\ref{subsec:spst}, we examine different arrangement of the hopfion chains and construct the stable 3D hopfion superstructure. In Sec.~\ref{subsec:distribution}, we present the detailed magnetic and topological properties of the 3D superstructure in both real and momentum space. In Sec.~\ref{subsec:metastability}, we evaluate the stability of the hopfion superstructure while varying the initial hopfion density. Section~\ref{sec:sum} is devoted to the summary of this study and perspectives.

%%%%%%%%%%%%%%%%%%%%%%%%%%%%%%%%%%%%%%%%%%%%%%%%%%%%%%%%%%%%%%%%%%%%%%%%%%%%%%%%%%%%%%%%%%%%%%%%%%%%%%%%%%%%%%%%%%%%%%%%%%%%%%%%%%%%

\section{Model and method}
\label{sec:model}
In this study, we consider a frustrated spin model with single-ion magnetic anisotropy on a simple cubic lattice. The Hamiltonian is given by
\begin{align}
\mathcal{H} = \sum_{\alpha = 1}^4 \sum_{\langle i,j \rangle_{\alpha}} J_{\alpha}~\bm{S}_i \cdot \bm{S}_j + K \sum_{i} (S_i^z)^2,
\label{eq:H}
\end{align}
where $\bm{S}_i = (S_i^x,S_i^y,S_i^z)$ represents the classical spin at site $i$ with $|\bm{S}_i|=1$. The first term represents the exchange interactions up to the fourth-neighbor sites on the cubic lattice; the sum $\langle i,j \rangle_{\alpha}$ runs over $\alpha$th-neighbor pairs. We set the coupling constants $J_1 = -1$, $J_2 = 0.1654$, $J_3 = 0$, and $J_4 = 0.0827$. We note that similar parameters were used for stabilizing a single hopfion~\cite{Bogolubsky1988,Rybakov2022}. The second term denotes the single-ion anisotropy; positive (negative) $K$ prefers in-plane (out-of-plane) spin configurations. We set the lattice constant as unity and adopt periodic boundary conditions.

We study stable spin configurations for the model in Eq.~\eqref{eq:H} by performing large-scale energy optimization based on the gradient descent. In the calculations, we first prepare an initial spin configuration $\{\bm{S}_i(\theta_i^{\rm{init}},\varphi_i^{\rm{init}})\}$, where $\theta_i^{\rm init}$ and $\varphi_i^{\rm init}$ denote the initial values of the polar and azimuthal angles of spin $\bm{S}_i$, respectively. Next, we compute the energy $\mathcal{H}$ and its gradient $\nabla \mathcal{H}$ with respect to the angle variables. Then, we obtain a new spin configuration $\{\bm{S}_i(\theta_i^{\rm{new}},\varphi_i^{\rm{new}})\}$ by updating all the angle variables along the gradient to reduce the energy. By repeating this procedure until the convergence of the energy and its gradient, we finally obtain an energetically stable spin configuration $\{\bm{S}_i(\theta_i^{\rm{final}},\varphi_i^{\rm{final}})\}$. We calculate the gradient by automatic differentiation and perform the optimization by Adam~\cite{Kingma2014}, exploiting JAX~\cite{jax2018github} and Optax libraries~\cite{deepmind2020jax}. This method is powerful and allows handling more than $10^6$ spins in a reasonable computational time. We set the learning rate in Adam to $0.01$. 

As our study focuses on magnetic hopfions, we perform the above optimization starting from initial spin configurations that include hopfions. We note that the model in Eq.~\eqref{eq:H} with the current parameter set exhibits a ferromagnetic order in the ground state. Nonetheless, our aim is to find spin configurations consisting of magnetic hopfions at local energy minima that serve as, at least, metastable states, remaining stable against perturbations. From this construction, we will also discuss the possible scenario for stabilizing such superstructures as the ground state in Sec.~\ref{sec:sum}.

In order to prepare an initial state including magnetic hopfions, we make use of the following ansatz describing a single hopfion with Hopf number $H = +1$~\cite{Wang2019}:
\begin{align}
\begin{split}
S^x_{i} &= \frac{4[2x_{i}z_{i} - y_{i}(r_i^2-1)]}{(1 + r_i^2)^2},\\
S^y_{i} &= \frac{4[2y_{i}z_{i} - x_{i}(r_i^2-1)]}{(1 + r_i^2)^2},\\
S^z_{i} &= 1 - \frac{8(x_{i}^2 + y_{i}^2)}{(1 + r_i^2)^2},
\label{eq:ansatz}
\end{split}
\end{align}
where $\bm{r}_i = (x_i, y_i, z_i)$ denotes the position of spin $\bm{S}_i$, and $r_i^2 = x_i^2 + y_i^2 + z_i^2$. The size of the hopfion can be varied by
\begin{align}
\begin{split}
r_i^{xy} &\rightarrow r_i^{xy \prime} = \frac{e^{r_i^{xy}/w_R}-1}{e^{R/w_R}-1}\\
z_{i} &\rightarrow z_{i}^{\prime} = \frac{z_{i}}{|z_{i}|}\frac{e^{|z_{i}|/w_h}-1}{e^{h/w_h}-1},
\label{eq:rescale}
\end{split}
\end{align}
where $r_i^{xy} = \sqrt{x_{i}^2 + y_{i}^2}$; $R$, $w_R$, $h$, and $w_h$ are the characteristic length scales parametrizing the shape of the hopfion. The mirror reflection while keeping the spin direction, e.g., $(x, y, z) \to (x, -y, z)$, reverses the Hopf number as $H = -1$. To prepare an initial state with multiple hopfions, we first optimize the spin configuration given by Eq.~\eqref{eq:ansatz} by using the above scheme, and then copy it with spatial translation and rotation. However, as the vector summation of $\bm{S}_i$ for $N_{\rm{hopf}}$ hopfions leads to $\bm{S}_i \to (0,0,N_{\rm{hopf}})$ for $\bm{r}_i$ far away from all the hopfion cores, we uniformly subtract a vector (0, 0, $N_{\rm{hopf}} - 1$) and normalize the length of all spins. In short, we prepare an initial state with $N_{\rm{hopf}}$ hopfions by superposing the spins $\bm{S}_i^j$ for $j$th hopfion generated by Eq.~\eqref{eq:ansatz} with appropriate translation and rotation, using
\begin{align}
\bm{S}_i^{\rm{init}} = \mathcal{N}_i \left(\sum_{j=1}^{N_{\rm{hopf}}}\bm{S}_i^j - (N_{\rm{hopf}}-1)\bm{\hat{z}}_i\right),
\end{align}
where $\mathcal{N}_i$ denotes the normalization at each site, and $\bm{\hat{z}}_i = (0, 0, 1)$. 

To characterize the topological property of the spin configurations obtained by the optimization scheme, we compute the Hopf number. The integral form of the Hopf number in continuum limit is defined by~\cite{Whitehead1947,Vega1978}
\begin{align}
H = -\int \bm{B}(\bm{r}) \cdot \bm{A}(\bm{r}) ~d\bm{r},
\label{eq:Hreal}
\end{align}
where $\bm{B}(\bm{r})$ is the emergent magnetic field whose $\alpha$ component is given by
\begin{align}
B^\alpha(\bm{r}) =\frac{1}{8\pi} \varepsilon_{\alpha\beta\gamma}\bm{S}(\bm{r}) \cdot \left\{ \nabla_{\beta} \bm{S}(\bm{r}) \times \nabla_{\gamma} \bm{S}(\bm{r})
\right\},
\label{eq:B}
\end{align}
and $\bm{A}(\bm{r})$ represents the vector potential satisfying $\nabla \times \bm{A}(\bm{r}) = \bm{B}(\bm{r})$; $\varepsilon_{\alpha\beta\gamma}$ represents the Levi-Civita symbol, and $\bm{S}(\bm{r})$ denotes the spin at position $\bm{r}$.  By performing Fourier transformation $\bm{B}(\bm{r}) = \frac1N \sum_{\bm{k}} \bm{B}(\bm{k}) \exp(i2\pi\bm{k} \cdot \bm{r})$ and $\bm{A}(\bm{r}) = \frac1N \sum_{\bm{k}} \bm{A}(\bm{k}) \exp(i2\pi\bm{k} \cdot \bm{r})$ and fixing gauge as $\bm{k} \cdot \bm{A} = 0$, which leads to
\begin{align}
    \bm{A}(\bm{k}) = -i\frac{\bm{k} \times \bm{B}(\bm{k})}{2\pi \bm{k}^2},
    \label{eq:vecpotential}
\end{align}
we can compute the Hopf number in momentum space rather than real space as~\cite{Moore2008,Liu2018}
\begin{align}
H = i\frac{1}{N}\sum_{\bm{k}} \frac{\bm{B}(-\bm{k}) \cdot [\bm{k} \times \bm{B}(\bm{k})]}{2\pi \bm{k}^2},
\label{eq:Hfourier}
\end{align}
where $N$ is the system size. In addition, to study the real-space distribution of the Hopf number, we also compute the Hopf number density defined by the integrand in Eq.~\eqref{eq:Hreal} as 
\begin{align}
\rho_{\rm H}(\bm{r}) = -\bm{B}(\bm{r}) \cdot \bm{A}(\bm{r}).
\end{align}

In the calculations for the model in Eq.~\eqref{eq:H}, we replace the integral in Eq.~\eqref{eq:Hreal} by a summation over the discrete lattice. Furthermore, we compute $\bm{B}(\bm{r})$ in Eq.~\eqref{eq:B} by a solid angle spanned by three spins as 
\begin{align}
B_i^{\alpha} = \frac{1}{8\pi}\varepsilon_{\alpha\beta\gamma}[\Omega(\bm{S}_i, &\bm{S}_{i+\beta}, \bm{S}_{i+\gamma}) \nonumber \\
& + \Omega(\bm{S}_{i+\gamma}, \bm{S}_{i+\beta}, \bm{S}_{i+\beta+\gamma})],
\label{eq:solidangle}
\end{align}
where $i+\beta$ denotes the neighboring site to site $i$ in the $\beta$ direction; $\Omega$ is the solid angle calculated by
\begin{align}
&\Omega(\bm{S}_i,\bm{S}_j,\bm{S}_k) \nonumber \\
& \qquad = 
2\arctan\left\{
\frac{\bm{S}_i \cdot (\bm{S}_j \times \bm{S}_k)}{1 + \bm{S}_i\cdot\bm{S}_j + \bm{S}_j\cdot\bm{S}_k + \bm{S}_k\cdot\bm{S}_i}
\right\},
\end{align}
where the arctangent on the right-hand side is implemented by the atan2 function to satisfy $\Omega \in [-2\pi, 2\pi)$. By using inverse-Fourier transformation of Eqs.~\eqref{eq:vecpotential} and \eqref{eq:solidangle}, we can obtain the vector potential in the discrete lattice system, and hence, compute the Hopf number density $\rho_{\rm H}(\bm{r}_i)$ at site $i$.

Besides the Hopf number and its density, we also study the skyrmion number $N_{\mathrm{sk}}$ on a 2D cut of the 3D spin structure, which is given by~\cite{Berg1981}
\begin{align}
    N_{\mathrm{sk}} = \sum_{i \in S_{\alpha}^{\perp}} B^{\alpha}_{i},
\label{eq:Nsk}
\end{align}
where the summation is taken over a 2D region $S_{\alpha}^{\perp}$ perpendicular to the $\alpha$ axis. We also compute the local scalar spin chirality $\chi_i^{\alpha}$
defined by
\begin{align}
\chi_i^{\alpha} = \frac{1}{2} \varepsilon_{\alpha\beta\gamma}[\bm{S}_i \cdot &(\bm{S}_{i+\beta} \times \bm{S}_{i+\gamma}) \nonumber \\
&+ \bm{S}_{i+\gamma} \cdot (\bm{S}_{i+\beta} \times \bm{S}_{i+\beta+\gamma})].
\end{align}
In addition, to characterize the spin configurations, we also compute the spin structure factor given by
\begin{align}
    S^{\alpha}(\bm{q}) = \frac{1}{N} \sum_{l,l^\prime} S^{\alpha}_{l} S^{\alpha}_{l^\prime} e^{-i \bm{q} \cdot (\bm{r}_{l} - \bm{r}_{l^\prime})}.
\label{eq:ssf}
\end{align}

%%%%%%%%%%%%%%%%%%%%%%%%%%%%%%%%%%%%%%%%%%%%%%%%%%%%%%%%%%%%%%%%%%%%%%%%%%%%%%%%%%%%%%%%%%%%%%%%%%%%%%%%%%%%%%%%%%%%%%%%%%%%%%%%%%%%
\section{Results}
\label{sec:results} 
In this section, we show the numerical results for the model in Eq.~\eqref{eq:H} by the energy optimization. First, we examine the interactions between two hopfions for various relative configurations in the isotropic case with $K = 0$ in Sec.~\ref{subsec:int}. We empirically find a ``color rule" for visually understanding the effective interactions between horizontally placed hopfions, and show that the hopfions can be stacked to form a 1D hopfion chain. Next, in Sec.~\ref{subsec:anisotropy}, we demonstrate that the hopfion chain becomes robust against perturbations by introducing the easy-plane anisotropy $K>0$. In Sec.~\ref{subsec:spst}, assembling the hopfion chains based on the color rule, we construct a 3D superstructure of hopfions. In Sec.~\ref{subsec:distribution}, we discuss the magnetic and topological properties of the hopfion superstructure. Finally, in Sec.~\ref{subsec:metastability}, we discuss the stability of the superstructure while varying the itinial hopfion density.

\subsection{Effective interactions between hopfions}
\label{subsec:int} 

\begin{figure}[tb]
    \begin{center}
    \includegraphics[width=\hsize]{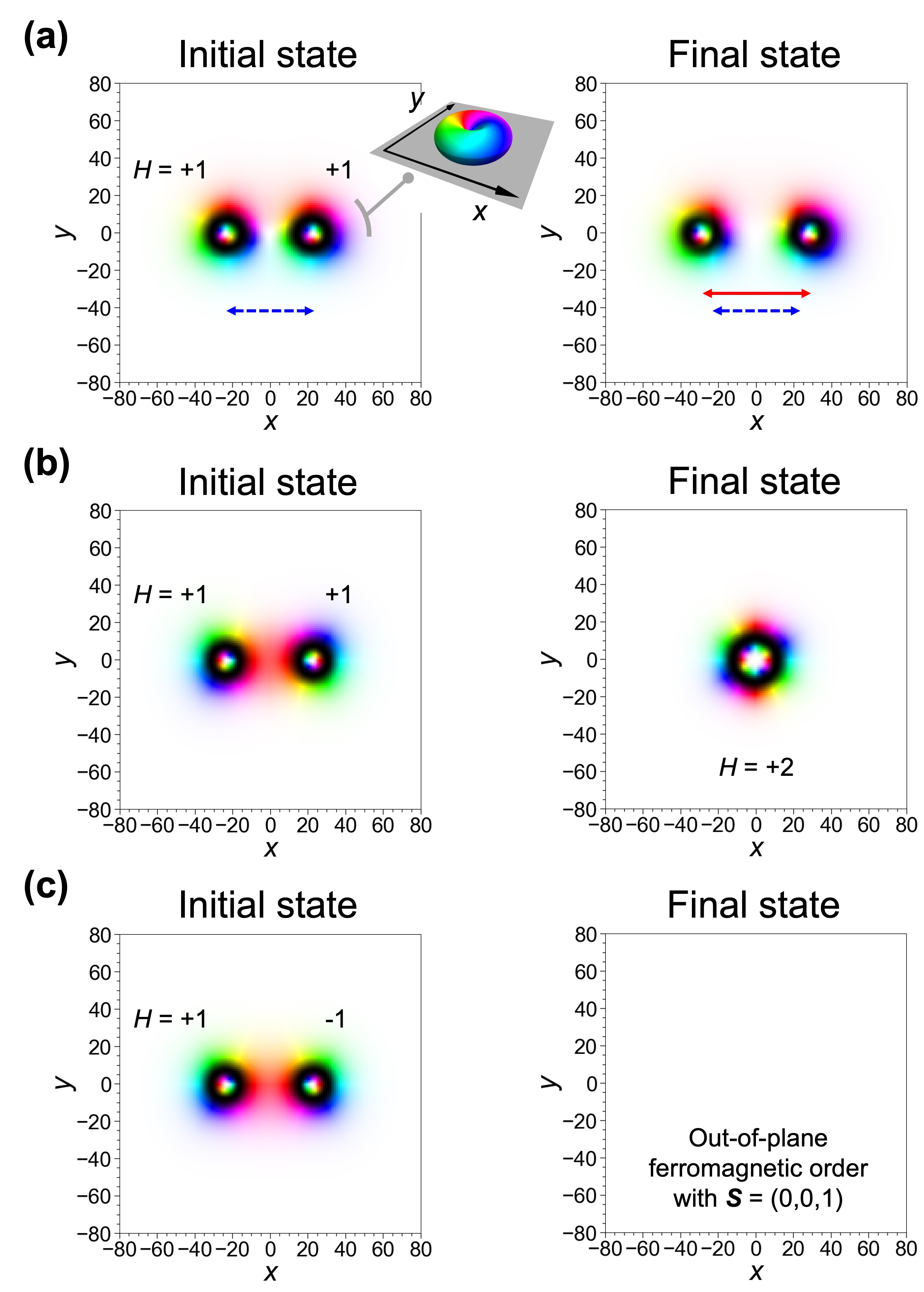}
    \caption{Energy optimization of two hopfions in horizontal arrangements. The left and right panels in each figure represent the spin configurations on the $xy$ cut through the hopfion cores in the initial and final states, respectively. The system size is $N=161^3$, and the final state is obtained after $10^5$ optimization steps. In (a), when two hopfions with Hopf number $H$ = +1 are prepared so that the spins in the proximate peripheries are antiparallel (complementary colors), they move away from each other after the energy optimization, suggesting a repulsive interaction between them. The blue dashed and red solid arrows indicate the distance between hopfions in the initial and final states, respectively. The color represents the spin direction as shown in the right panel of Fig.~\ref{fig:hierarchy}(a). In (b), the two hopfions in the initial state are rotated in the opposite directions so that the proximate spins are parallel to each other (same colors). After the optimization, two hopfions cause a fusion into one with a higher Hopf number $H = +2$, suggesting an attractive interaction. In (c), a similar initial state with $H = +1$ and $-1$ causes pair annihilation, suggesting again an attractive interaction. 
    }
    \label{fig:horizontal}
    \end{center}
\end{figure}

To construct an energetically-stable superstructure of periodically arranged hopfions, it is crucial to clarify the fundamental interactions between hopfions. The hopfion has a 3D donut-like shape, and hence, the interactions may depend on the relative configurations of two hopfions. Such interactions were partly studied for the present model~\cite{Rybakov2022} and the Skyrme-Faddeev (SF) model~\cite{Ward2000,Hietarinta2012}. 
Here, we perform comprehensive studies for various relative configurations and Hopf numbers. In this subsection, we consider the isotropic model with $K = 0$ in Eq.~\eqref{eq:H}.

Let us first discuss the cases starting from the initial states in which two hopfions are arranged horizontally, i.e., the two hopfions are placed on the same plane. Figure~\ref{fig:horizontal}(a) shows the results of energy optimization starting from two hopfions with $H=+1$ each in such a horizontal arrangement. In this case, we prepare the initial state by placing two copies of a single hopfion apart, and hence, the spins on the proximate peripheries of two hopfions are antiparallel to each other, as indicated by complementary colors, purple and green. After the optimization, the two hopfions move away from each other, suggesting a repulsive interaction between them. In contrast, when the two hopfions in the initial state are rotated so that the proximate spins are parallel (same colors), they move toward each other, suggesting an attractive interaction. For a pair with $H=+1$ [Fig.~\ref{fig:horizontal}(b)], the two hopfions cause a fusion into one with a higher Hopf number $H=+2$, while for a pair with $H=+1$ and $H=-1$ [Fig.~\ref{fig:horizontal}(c)], two hopfions cause pair annihilation and disappear. We note that a similar fusion was reported in the previous studies~\cite{Ward2000, Hietarinta2012,Rybakov2022}. In particular, it was shown in Ref.~\cite{Rybakov2022} that the energy of a hopfion with Hopf number $H$ is proportional to $H^{3/4}$, indicating that a single hopfion with Hopf number $2H$ has a lower energy than two independent hopfions with Hopf number $H$. The result for the fusion in Fig.~\ref{fig:horizontal}(b) looks consistent with this statement, but that in Fig.~\ref{fig:horizontal}(a) suggests that the instability depends on the relative configurations of hopfions. Thus, our results indicate that, in the horizontal arrangements, there is an energy barrier for two hopfions with antiparallel spins in the proximate peripheries, whereas no barrier for those with parallel spins. This is simply summarized as the color rule: When two hopfions face complementary colors each other, they feel a repulsive force and remain stable, while when they face the same color, they feel an attractive force and cause fusion or pair annihilation.

\begin{figure}[t!]
    \begin{center}
    \includegraphics[width=\hsize]{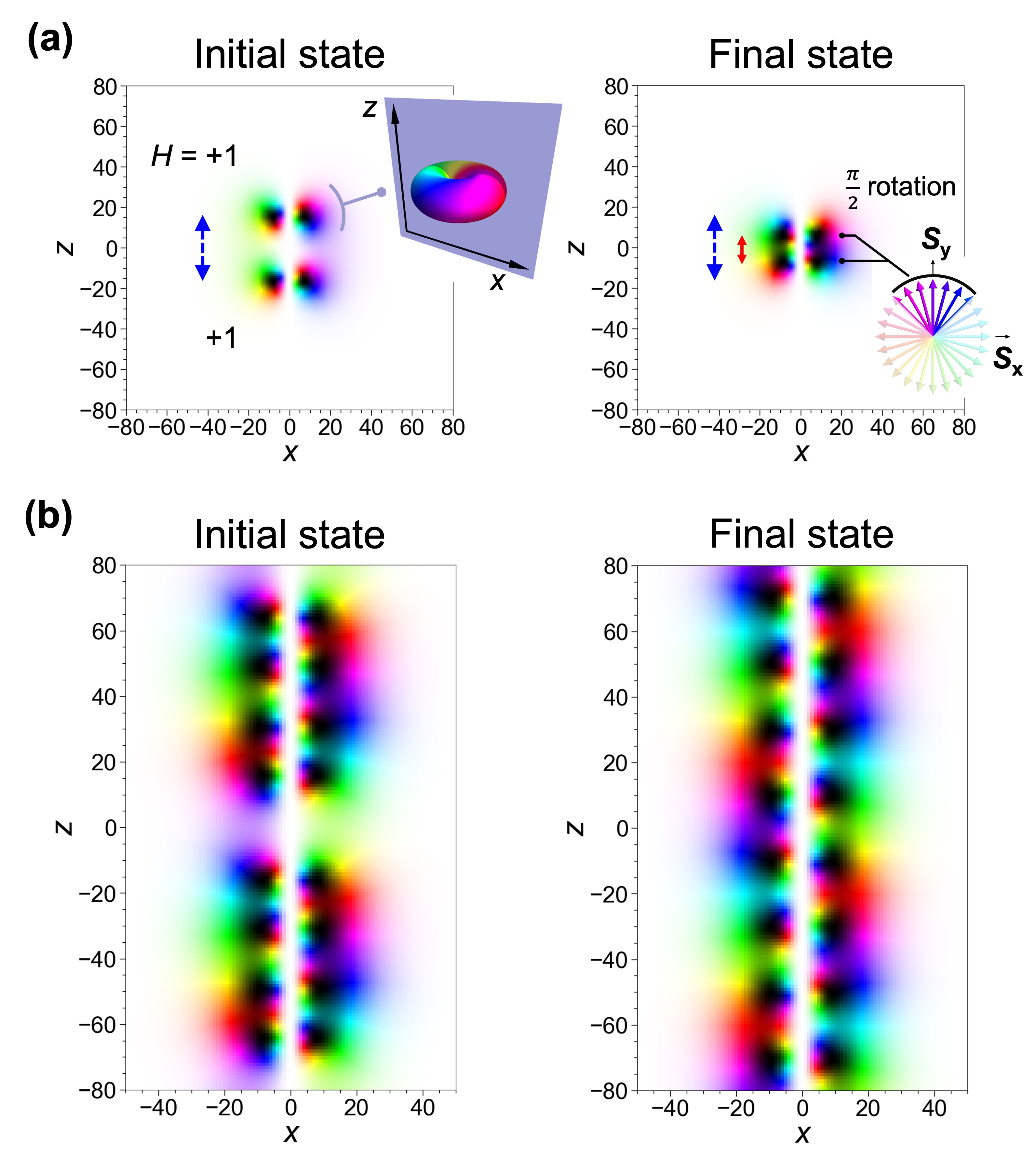}
    \caption{Energy optimization of two hopfions in vertical arrangements. Each panel represents the spin configurations on the $xz$ cut through the hopfion cores. The notations are common to those in Fig.~\ref{fig:horizontal}. In (a), when two hopfions with Hopf number $H$ = +1 are prepared with a spatial translation in the $z$ direction, they approach each other and stay at a distance without fusion after the energy optimization, suggesting a short-range repulsive interaction in addition to an overall attractive one. The inset in the right panel explicitly shows the relative rotation between two hopfions. In (b), the initial state is prepared by adding two hopfions to the initial spin state of (a), creating a chain of four hopfions through energy optimization up to $1.5 \times 10^5$ steps, copying it, and arranging them vertically. After the optimization, the eight hopfions align periodically by adjusting the distances from each other. The system size is $N=161^3$ for (a) and $N=101\times 101\times 161$ for (b), and the optimization is performed up to $10^5$ steps in both cases.
    }
    \label{fig:vertical}
    \end{center}
\end{figure}

Next, we discuss the cases starting from the initial states in which the two hopfions are placed in a vertical arrangement. Figure~\ref{fig:vertical}(a) shows the results of energy optimization starting from two hopfions with $H$ = +1 each. They approach each other through the optimization, suggesting an attractive interaction. However, in contrast to Fig.~\ref{fig:horizontal}(b), the two hopfions do not cause fusion, but stay at a distance to form a bound state. The total Hopf number is kept at $+2$. A similar bound state was discussed for the SF model~\cite{Ward2000}. The results suggest a short-range repulsive interaction between the hopfions in addition to the long-range attractive one. It is worthy noting that the two hopfions in the final state are rotated to each other around the $z$ axis by approximately $\pi/2$. We find that the initial states with different relative rotation angles between the two hopfions always converge to this rotated state (not shown). We note that the color rule discussed above is not straightforwardly applied to the vertical arrangement.

Given the stable bound state in the vertical arrangement, we try a vertical stacking of multiple hopfions. Figure~\ref{fig:vertical}(b) shows an example of eight hopfions. In this case, we prepare the initial state by first adding one hopfion to top and bottom of the initial state in Fig.~\ref{fig:vertical}(a), next optimizing the spin configurations of the four hopfions, and finally copying the result with a spatial translation. The energy optimization starting from this initial state gives the stable chain of the eight hopfions, as shown in the right panel of Fig.~\ref{fig:vertical}(b). Note that the procedure of the initial state preparation indicates that a chain of four hopfions is also stable. In both cases of four and eight hopfions, the neighboring hopfions are rotated to each other by approximately $\pi/2$ as in the two-hopfion bound state in Fig.~\ref{fig:vertical}(a). The results imply that vertical chains of multiple hopfions with a twist of $\sim \pi/2$, i.e., with period four, are energetically stable. 

%%%%%%%%%%%%%%%%%%%%%%%%%%%%%%%%%%%%%%%%%%%%%%%%%%%%%%%%%%%%%%%%%%%%%%%%%%%%%%%%%%%%%%%%%%%%%%%%%%%%%%%%%%%%%%%%%%%%%%%%%%%%%%%%%%%%
\subsection{Effect of easy-plane anisotropy}
\label{subsec:anisotropy}

\begin{figure*}[t!]
    \begin{center}
    \includegraphics[width=\hsize]{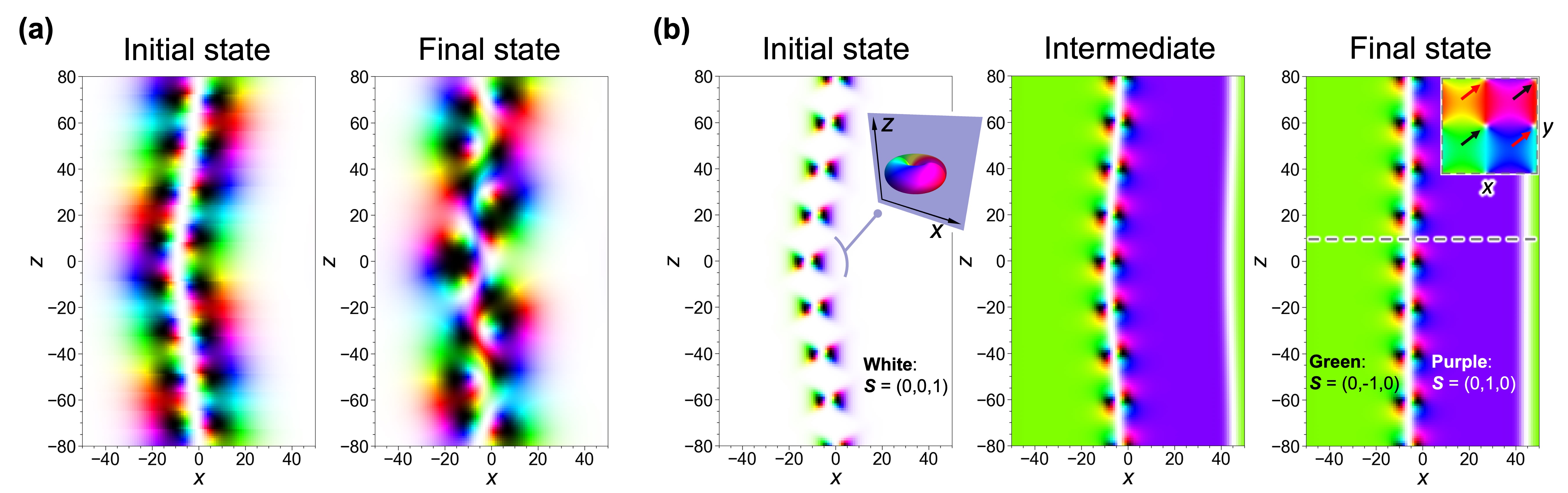}
    \caption{Stabilization of a hopfion chain by the easy-plane anisotropy. (a) In the absence of the easy-plane anisotropy ($K=0$), the hopfion chain prepared with small warping is unstable and disordered after the optimization. (b) In the presence of the easy-plane anisotropy with $K = 0.01$, such warping is straightened up through the optimization, indicating that the easy-plane anisotropy stabilizes the hopfion chain against the perturbation. In the final state, the spins surrounding the hopfion chain are aligned in-plane by the easy-plane anisotropy. The inset displays the spin configuration on the $xy$ plane at $z = 10$ represented by the gray dashed line in the main panel. It accommodates a magnetic meron around the origin, and in addition, one meron and two antimerons near the boundary of the system so as to compromise the in-plane spin configuration under the periodic boundary conditions; the black and red arrows denote the merons with $N_{\rm sk}=+1/2$ and the antimerons with $N_{\rm sk}=-1/2$. The system size is $N=101\times 101\times 161$, and the optimization is performed up to $10^5$ and $5 \times 10^4$ steps in (a) and (b), respectively. The intermediate state in (b) is taken from $10^4$ step during the optimization.}
    \label{fig:easy-plane}
    \end{center}
\end{figure*}

Above we show stable chains of multiple hopfions. However, they are obtained by the optimization starting from the initial states with perfect vertical alignment of the hopfions, and might be unstable against a misalignment. We examine such an instability by preparing an initial state with small warping along the chain, as shown in the left panel of Fig.~\ref{fig:easy-plane}(a). Indeed, we find that the hopfion chain is unstable and disordered after the optimization as shown in the right panel, while the total Hopf number is preserved around $+8$. This indicates that the hopfion chains are at a saddle point on the energy surface. A similar difficulty in aligning hopfions has been suggested for the SF model~\cite{Hietarinta2012}.

To circumvent the instability, we introduce the easy-plane single-ion anisotropy $K>0$ in Eq.~\eqref{eq:H}. Figure~\ref{fig:easy-plane}(b) shows the energy optimization with $K=0.01$, starting from the initial state prepared by the ansatz in Eq.~\eqref{eq:ansatz} with small warping. In the early stage of the optimization, the spins surrounding the hopfion chain are aligned in the $xy$ plane due to the in-plane anisotropy, while the spins inside the hopfion chain remain intact to form an up-spin chain penetrating the hopfion cores. During the optimization, the hopfion chain is gradually straightened up, as shown in the middle panel of Fig.~\ref{fig:easy-plane}(b), and eventually resulting in the final state with perfect alignment in the right panel. Thus, the multiple-hopfion chain is stabilized by the easy-plane anisotropy against perturbations.

It is worthy noting that, in the optimized state under the easy-plane anisotropy, the hopfion chain is not twisted, unlike in Fig.~\ref{fig:vertical}(b). In addition, since it is surrounded by spins lying on the $xy$ plane that rotate around the chain, the 2D cut between two neighboring hopfions exhibits a topological texture regarded as a magnetic meron with skyrmion number $N_{\rm sk}=1/2$; see the inset of the right panel of Fig.~\ref{fig:easy-plane}(b). In this case, there appear additionally one meron and two antimerons near the boundary of the system so as to compromise the in-plane spin configuration under the periodic boundary conditions; the two merons with $N_{\rm sk}=+1/2$ and two antimerons with $N_{\rm sk}=-1/2$ are located in a staggered manner to form a square lattice. The result suggests that, in the presence of the easy-plane anisotropy, a string of magnetic merons nested with hopfions is formed, contributing to the stability to the hopfion chain. We note that, in contrast, the easy-axis anisotropy with $K < 0$ does not stabilize such a meron string, and hence a hopfion chain, since it does not lay the background spins into the $xy$ plane.

%%%%%%%%%%%%%%%%%%%%%%%%%%%%%%%%%%%%%%%%%%%%%%%%%%%%%%%%%%%%%%%%%%%%%%%%%%%%%%%%%%%%%%%%%%%%%%%%%%%%%%%%%%%%%%%%%%%%%%%%%%%%%%%%%%%%
\subsection{Stable 3D hopfion superstructure}
\label{subsec:spst}

\begin{figure*}[tb]
    \begin{center}
    \includegraphics[width=\hsize]{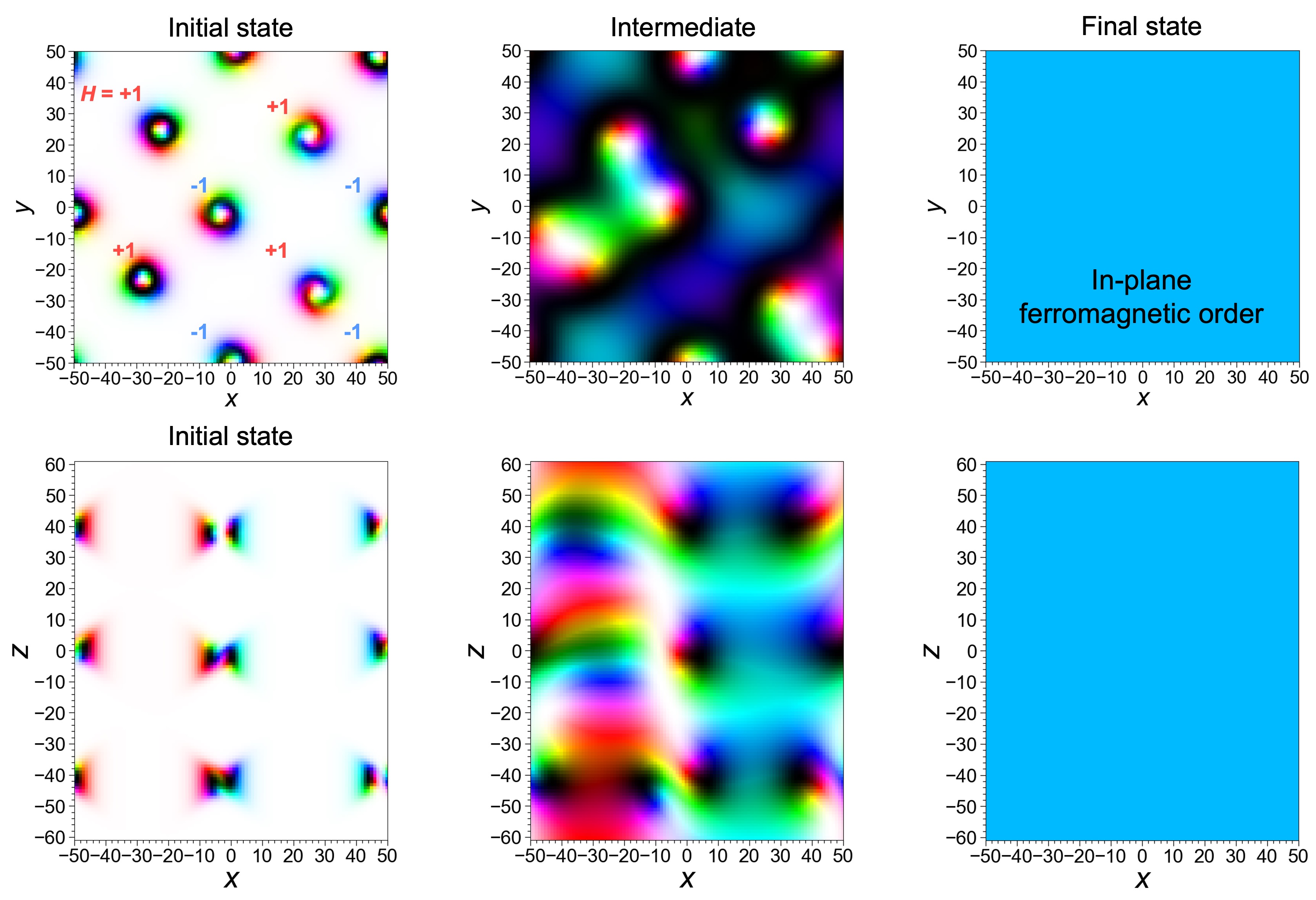}
    \caption{
    Instability of hopfion chains in a staggered arrangement. The upper and lower panels show the spin configurations in the $xy$ plane at $z=0$ and in the $xz$ plane at $y=0$, respectively. We take $K = 10^{-4}$ and $N=101\times 101\times 123$. The optimization is performed up to $3 \times 10^5$ steps, and the intermediate state in the middle column shows a snapshot at $10^4$ steps. In the initial state, each chain is composed of three hopfions prepared with small misalignment by random shifts, and the chains with Hopf number $H=+1$ and $-1$ are placed in a staggered manner to form a distorted square lattice in the $xy$ plane. During the optimization, the hopfion chains cause pair annihilation, resulting in an in-plane ferromagnetic order without hopfions whose moment direction in the $xy$ plane depends on the initial state. 
    }
    \label{fig:3D_noflip}
    \end{center}
\end{figure*}

Given the stable hopfion chain nested by merons, we here try to obtain a 3D hopfion superstructure composed of such hopfion chains. In Fig.~\ref{fig:easy-plane}(b), we found meron and antimeron strings forming a square lattice, one of which accommodates the hopfions but the others do not. This suggests the possibility of arranging the hopfion chains bound to meron and antimeron strings in a square-lattice manner. We first try an initial configuration by preparing a square lattice of hopfion chains with all $H=+1$, but fail to stabilize it; the energy optimization gives rise to additional meron and antimeron strings in the interstitials of the square lattice, which destabilizes the hopfion chains. Instead, we find that the hopfion chains remain stable when $H=+1$ and $-1$ are prepared in a staggered way, with a proper arrangement of spin directions. Below, we explain step-by-step how to stabilize the superstructure with staggered hopfion chains. 

Figure~\ref{fig:3D_noflip} shows a trial starting from the initial state prepared by placing hopfion chains with $H=\pm 1$ in a staggered manner to form a square lattice in the $xy$ plane. We take $K=10^{-4}$ here and hereafter. The hopfion chains are rotated to each other to exploit the knowledge of the color rule between neighboring hopfion chains that was found in Sec.~\ref{subsec:int}. To examine the stability against perturbations, all the hopfions are slightly shifted to random directions in the 3D space. Note that the period in the $z$ direction is about twice as long as that in Fig.~\ref{fig:easy-plane}(b) for the smaller $K$. Through the optimization, however, we find that hopfions move attractively and cause pair annihilation, resulting in the ground state with in-plane ferromagnetic order, as shown in the middle and right columns in Fig.~\ref{fig:3D_noflip}. Thus, a simple staggered arrangement of hopfion chains with $H=\pm 1$ is not stable.

\begin{figure*}[t!]
    \begin{center}
    \includegraphics[width=\hsize]{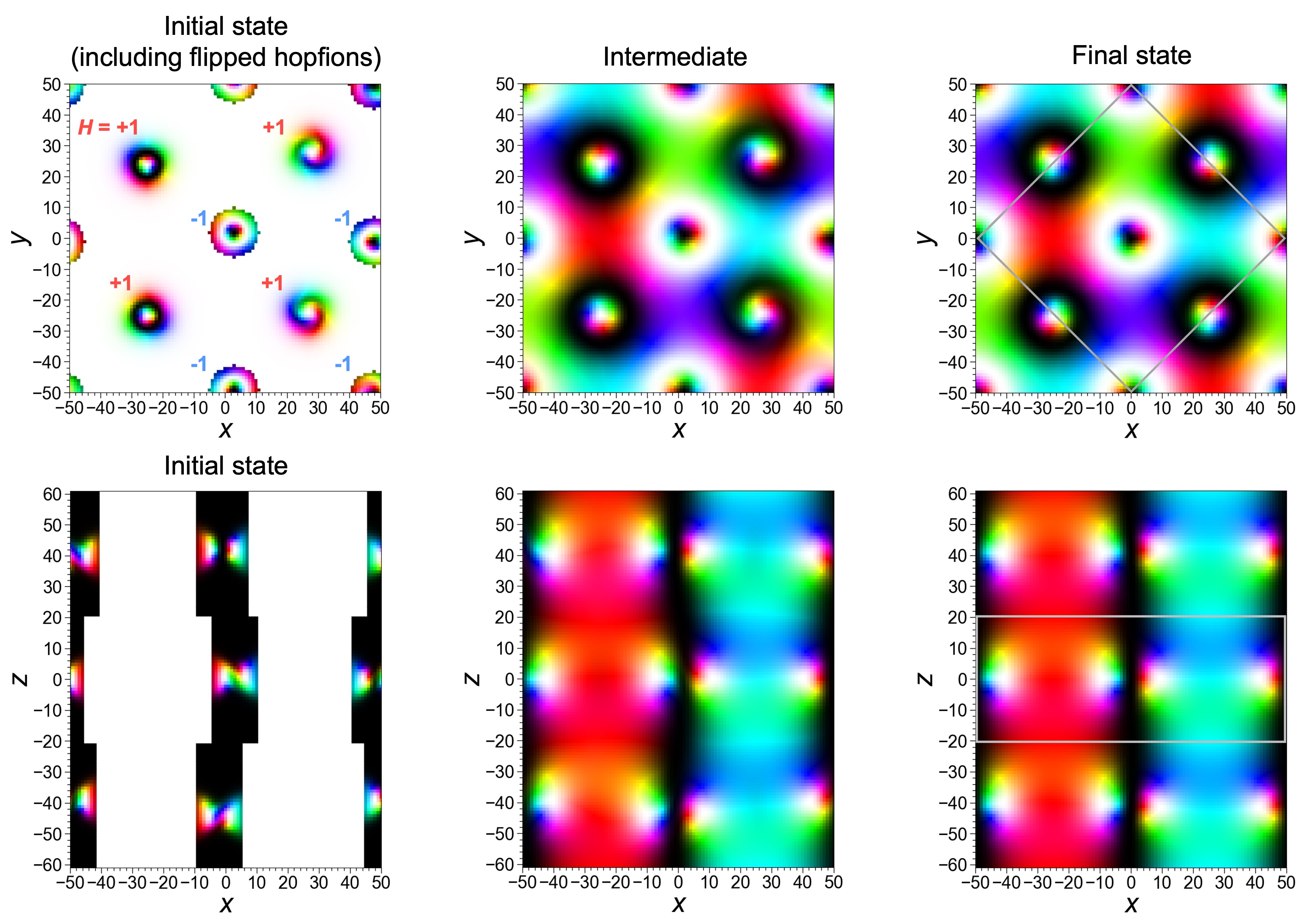}
    \caption{Stabilization of a 3D hopfion superstructure. The initial state is prepared by flipping the $z$ component of spins around the hopfion chains with $H=-1$ in the initial state in Fig.~\ref{fig:3D_noflip}; the spins surrounding the hopfion chains with $H=-1$ are flipped from up (white) to down (black), as shown in the lower-left panel. The other conditions are the same as Fig.~\ref{fig:3D_noflip}. Through the optimization, the hopfion chains do not cause pair annihilation, and they are straightened and aligned periodically to form a 3D superstructure of hopfions nested by magnetic merons. The gray square and rectangle in the right panels represent the magnetic unit cell. The bird’s-eye view of the final state is shown in Fig.~\ref{fig:hierarchy}(d). 
    }
    \label{fig:3D_flip}
    \end{center}
\end{figure*}

We find that such an instability can be circumvented by preparing the initial state by flipping the $z$ component of spins in the hopfion chains with $H = -1$. The left column of Fig.~\ref{fig:3D_flip} shows the initial state with spin-flipped hopfions. Note that the color rule between the neighboring hopfion chains is not hampered since only $S_i^z$ is flipped and the color at the periphery of each hopfion dose not change. In this case, in contrast to Fig.~\ref{fig:3D_noflip}, hopfions show no pair annihilation and form a stable 3D superstructure, as shown in the middle and right columns in Fig.~\ref{fig:3D_flip}. This suggests a repulsive interaction between the up-spin (white) and down-spin (black) hopfions in the $xy$ plane, in consistent with the color rule for complementary colors (antiparallel spins) found in Sec.~\ref{subsec:int}. In this way, we can successfully obtain a periodically-arranged 3D hopfion superstructure. This superstructure has a staggered arrangement of hopfions with $H=+1$ and up-spin cores and those with $H=-1$ and down-spin cores in the $xy$ slices, and between these slices, it has a staggered arrangement of magnetic merons whose cores are up and down spins. Note that in this case the 2D spin structure is composed of only merons, in contrast to merons and antimerons in Fig.~\ref{fig:easy-plane}(b); the hopfion chains with $H=+1$ ($-1$) are nested by merons with up(down)-spin cores. The 3D picture is shown in Fig.~\ref{fig:hierarchy}(d).

%%%%%%%%%%%%%%%%%%%%%%%%%%%%%%%%%%%%%%%%%%%%%%%%%%%%%%%%%%%%%%%%%%%%%%%%%%%%%%%%%%%%%%%%%%%%%%%%%%%%%%%%%%%%%%%%%%%%%%%%%%%%%%%%%%%%
\subsection{Magnetic and topological properties of hopfion superstructure}
\label{subsec:distribution}

\begin{figure}[tb]
    \begin{center}
    \includegraphics[width=\hsize]{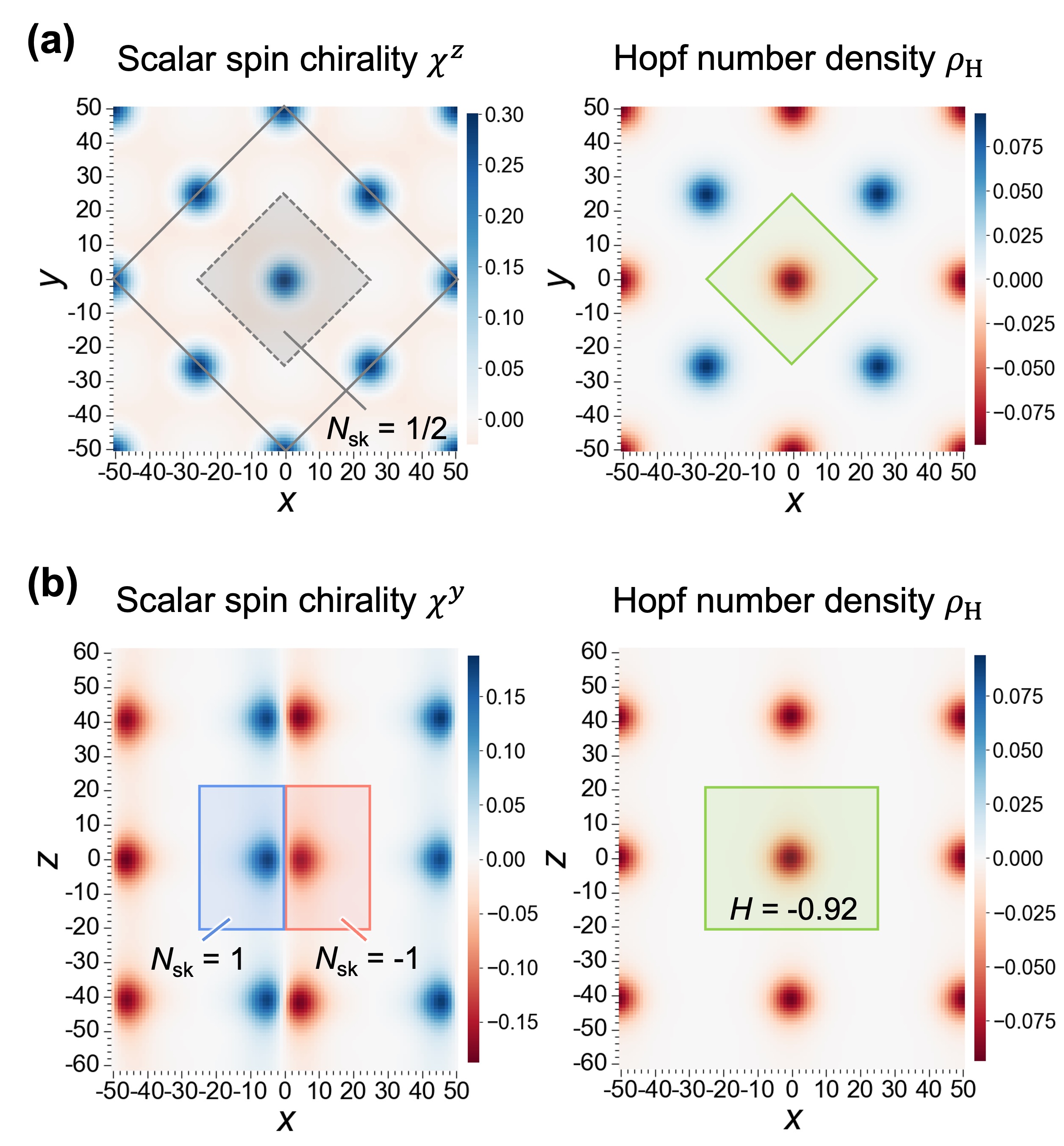}
    \caption{Real-space distributions of the local scalar spin chirality and the Hopf number density in the 3D hopfion superstructure obtained in Fig.~\ref{fig:3D_flip}: (a) and (b) are respectively the $xy$ and $xz$ cuts traversing the hopfion cores. In the left panels of (a) and (b), the $z$ and $y$ components of the local scalar spin chirality are respectively shown. The gray, red, and blue shaded areas represent the regions for computing the skyrmion number $N_{\mathrm{sk}}$, while the green is for the Hopf number $H$. The black square in the left panel of (a) denotes the 2D magnetic unit cell. See also Fig.~\ref{fig:hierarchy}(d).
    }
    \label{fig:toponum}
    \end{center}
\end{figure}

\begin{figure}[t!]
    \begin{center}
    \includegraphics[width=\hsize]{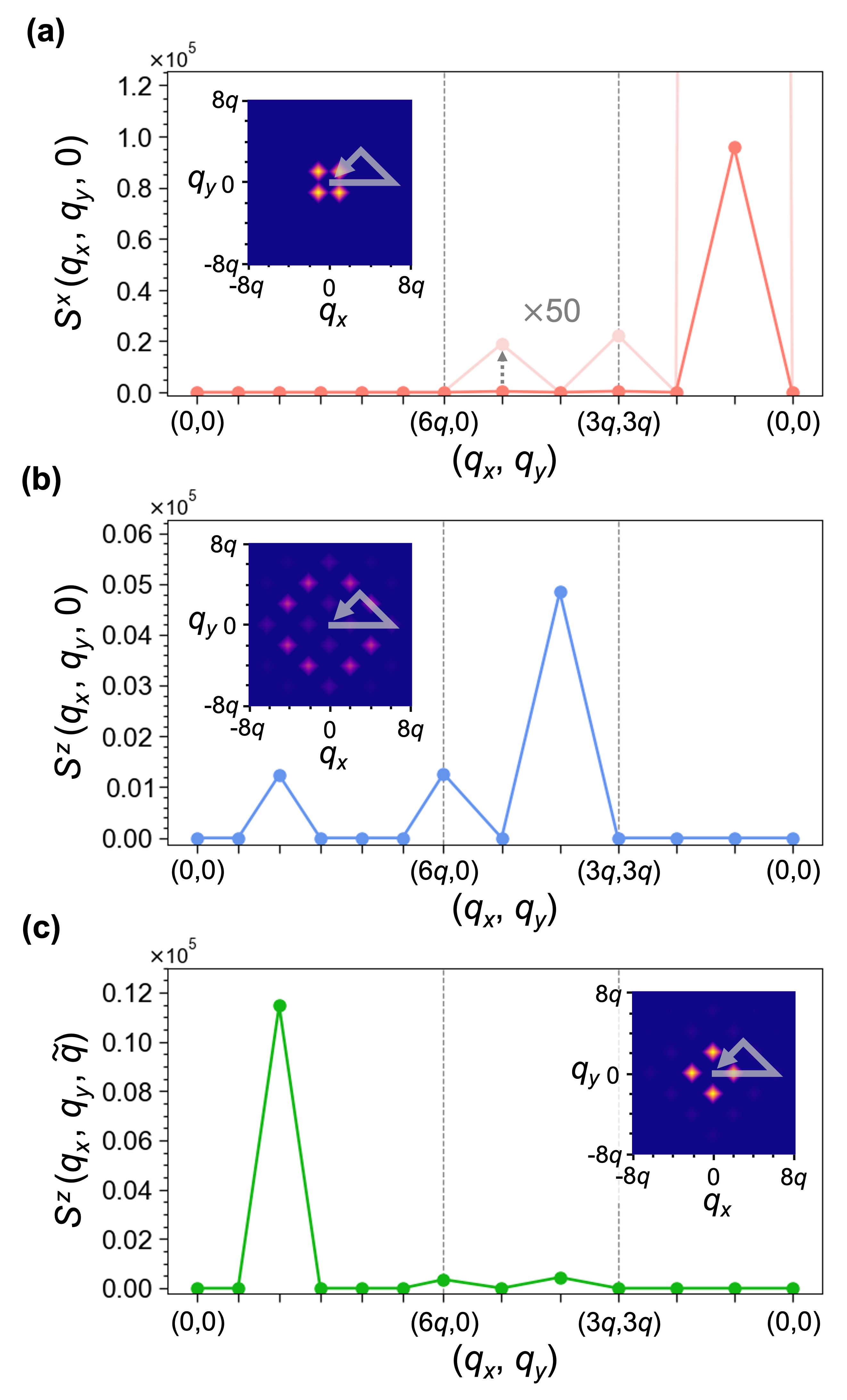}
    \caption{Spin structure factor of the 3D hopfion superstructure obtained in Fig.~\ref{fig:3D_flip}. (a) displays the $S^x$ component at $q_z$ = 0. The data multiplied by 50 are also presented in light red for better showing small additional peaks. (b) and (c) correspond to the $S^z$ components at $q_z$ = 0 and $q_z = \tilde{q} = 2\pi/41$, respectively. The insets in (a) and (b) are the 2D plots on the $q_z = 0$ plane, and that in (c) for the $q_z = \tilde{q}$ plane. The data in each panel are plotted along the gray arrows in the insets. The unit of momentum $q$ is set by the system size in the $x$ and $y$ directions as $q = 2\pi/101$.}
    \label{fig:ssf}
    \end{center}
\end{figure}

Figure~\ref{fig:toponum} shows the real-space distribution of the local scalar spin chirality and the Hopf number density on $xy$ and $xz$ cuts of the 3D hopfion superstructure obtained in Fig.~\ref{fig:3D_flip}. On the $xy$ cut plotted in Fig.~\ref{fig:toponum}(a), which traverses the hopfion cores in the hopfion square lattice, the $z$ component of the local scalar spin chirality takes a nonzero value periodically around the hopfion cores in the slightly negative background. The skyrmion number per hopfion is $N_{\rm sk}=1/2$ as shown in the left panel of Fig.~\ref{fig:toponum}(a); namely, $N_{\rm sk}=2$ per 2D magnetic unit cell. We note that the skyrmion number remains the same for all the $xy$ cuts irrespective of the $z$ coordinate. This indicates that all the 2D cuts are topologically equivalent to the meron lattice, and hence, the 3D hopfion superstructure can be viewed as a square lattice of meron strings running in the $z$ direction which thread the hopfion chains. A similar structure called the hopfion ring pierced by the skyrmion string has been reported~\cite{Zheng2023}. Meanwhile, the Hopf number density $\rho_{\rm H}(\bm{r}_i)$ appears with positive and negative values around the hopfion cores, as expected for the staggered arrangement of hopfion chains with $H = \pm1$. 

Figure~\ref{fig:toponum}(b) displays the $xz$ cut traversing the central axes of hopfion chains with $H=-1$. In the left panel, we find a periodic arrangement of pairs of positive (blue) and negative (red) values of the $y$ component of the local scalar spin chirality. This is consistent with the fact that each hopfion consists of a closed loop of a twisted skyrmion string; the skyrmion number integrated around each region is close to $N_{\rm sk}=+1$ and $-1$, as shown in the figure. Note that on a cut across the axes of hopfion chains with $H=+1$, the colors are inverted from Fig.~\ref{fig:toponum}(b), reflecting the reversal of the Hopf number. The Hopf number density in the right panel of Fig.~\ref{fig:toponum}(b) shows a negative value around the hopfion cores, as expected for the $H=-1$ hopfion chains. However, the Hopf number integrated over the green region in the right panels of Figs.~\ref{fig:toponum}(a) and \ref{fig:toponum}(b) is about $-0.92$, which is considerably deviates from $-1$. There are three possible reasons for this deviation: the dependence on the integrated area, the effect of lattice discretization in Eq.~\eqref{eq:Hreal}, and the fact that the background spins are not uniformly polarized unlike for an isolated hopfion. The former two appear to be irrelevant since we confirm that the Hopf number for an isolated single hopfion integrated in a similar area is close to $\pm 1$. Meanwhile, the last one might be the primary reason since it makes the twist of the skyrmion string ill-defined. Note that the Hopf number $H$ is expressed as $H = N_{\textrm{sk}} T$, where $T$ is the twist of the skyrmion string that forms the hopfion~\cite{Gladikowski1997}, and we obtain $|N_{\textrm{sk}}|$ close to 1, as shown in the left panel of Fig.~\ref{fig:toponum}(b).

Figure~\ref{fig:ssf} presents the spin structure factor in Eq.~\eqref{eq:ssf} of the 3D hopfion superstructure: (a) for the $x$ component, $S^{x}(q_x,q_y,0)$, which is equivalent to $S^{y}(q_x,q_y,0)$, (b) and (c) for the $z$ component, $S^{z}(q_x,q_y,0)$ and $S^{z}(q_x,q_y,\tilde{q})$, respectively, where $\tilde{q} = 2\pi/41$, along the gray lines in the Brillouin zone shown in each inset. In Fig.~\ref{fig:ssf}(a), $S^{x}(q_x,q_y,0)$ exhibits the largest peak at $\bm{q} = (q,q,0)$, while much smaller intensities are also present at the higher harmonics. These multiple higher harmonics are characteristic of multiple-$Q$ spin state composed of a superposition of multiple spin density waves. Similarly, $S^{z}(q_x,q_y,0)$ in Fig.~\ref{fig:ssf}(b) exhibits subdominant peaks with different intensities at different $\bm{q}$. For $q_z \neq 0$, no distinct intensities are observed for $S^{x}(\bm{q})$, whereas for $S^{z}(\bm{q})$ in Fig.~\ref{fig:ssf}(c), sharp peaks approximately $10$ times stronger than the subdominant peaks located in the $q_z = 0$ plane appear at $\bm{q} = (2q,0,\pm \tilde{q})$ and their $C_4$-symmetric positions. These peculiar features of the spin structure factor will be useful to identify the hopfion superstructure in experiments, such as neutron scattering.

%%%%%%%%%%%%%%%%%%%%%%%%%%%%%%%%%%%%%%%%%%%%%%%%%%%%%%%%%%%%%%%%%%%%%%%%%%%%%%%%%%%%%%%%%%%%%%%%%%%%%%%%%%%%%%%%%%%%%%%%%%%%%%%%%%%%
\subsection{Metastability of the hopfion superstructure}
\label{subsec:metastability}
\begin{figure}[t!]
    \begin{center}
    \includegraphics[width=\hsize]{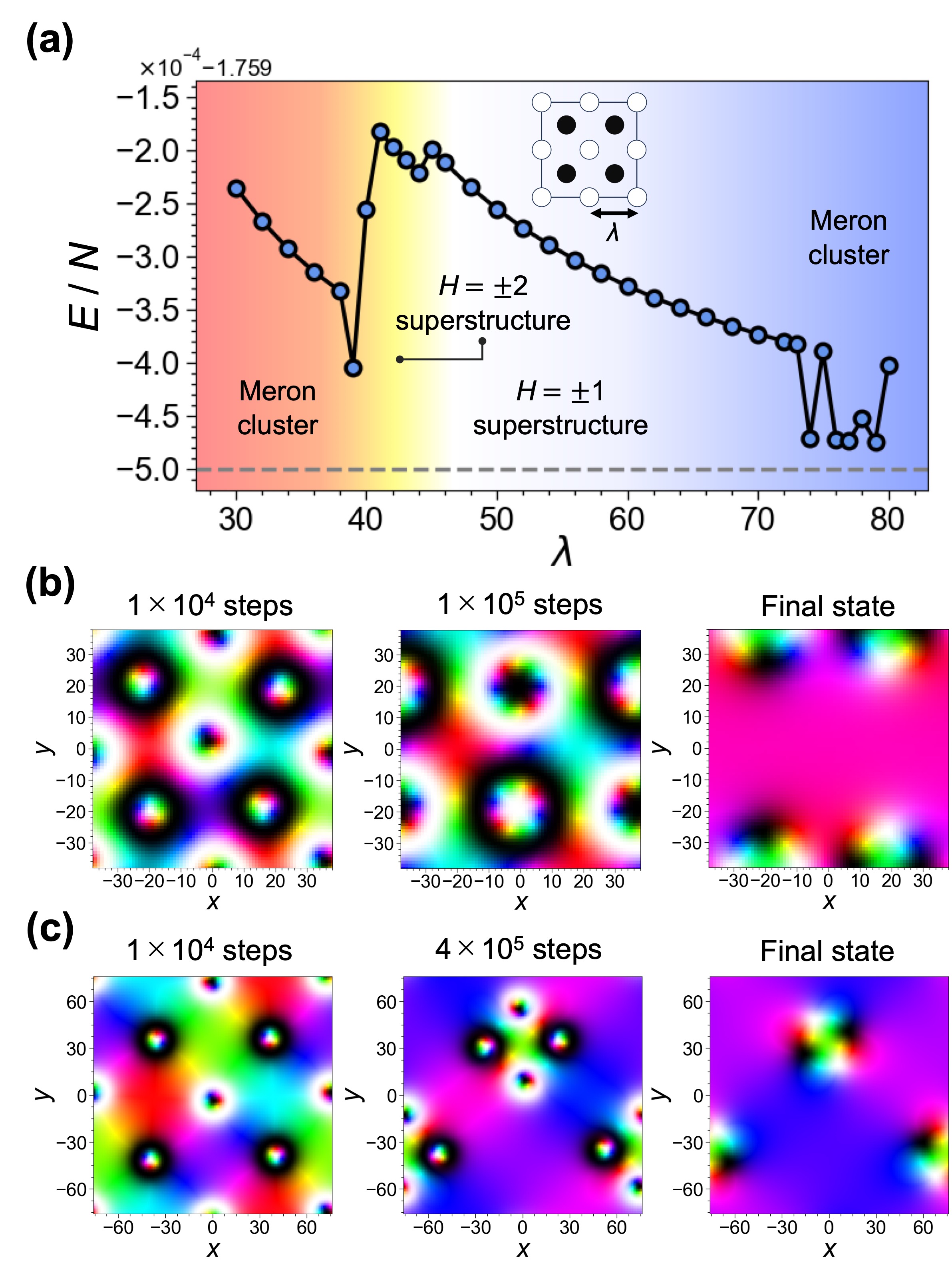}
    \caption{Metastability of the 3D hopfion superstructure. The optimization is performed up to $6 \times 10^5$ steps starting from a 2D arrangement of hopfions with period $\lambda$ in the system size $N = (2\lambda + 1)^2\times 41$ [see the inset of (a)]. (a) Optimized energy per spin (with the offset of $-1.759$) as a function of $\lambda$. The horizontal gray dashed line at $E/N = -1.7595$ corresponds to the energy of the ground state with in-plane ferromagnetic order. The superstructure composed of $H = \pm1$ survives as a metastable state in the white region of $45 \lesssim \lambda \lesssim 56$. In the yellow region of $41 \lesssim \lambda \lesssim 45$, we obtain another superstructure composed of $H = \pm2$. The hopfion superstructures become unstable in the other regions. See the text for the details. [(b),(c)] Instability of the hopfion superstructure through the optimization process for (b) $\lambda = 38$ and (c) $\lambda = 76$.}
    \label{fig:meta}
    \end{center}
\end{figure}

The hopfion superstructure obtained by the energy optimization is a metastable state at a local energy minimum, whose energy is slightly higher than that for the ferromagnetically ordered state, for the current parameter set. We here examine the metastability while varying the magnetic period in the $xy$ plane, $\lambda$. For this purpose, we perform the energy optimization in the system with $N=(2\lambda+1)^2 \times 41$ while keeping the total number of hopfions in the initial state as eight, consisting of four $H=+1$ hopfions and four $H = -1$ hopfions [see the inset of Fig.~\ref{fig:meta}(a)]. Note that, for simplicity,  the system size in the $z$ direction is set to one third of that in Fig.~\ref{fig:3D_flip}, corresponding to a single period of the superstructure in the $z$ direction.

Figure~\ref{fig:meta}(a) shows the energy per site after optimization as a function of $\lambda$. We find that the hopfion superstructure remains stable for $45 \lesssim \lambda \lesssim 56$ (white region), even after $6 \times 10^5$ optimization steps, twice as long as that in Fig.~\ref{fig:3D_flip}. Note that the energy difference with the ferromagnetic ground state is very small, of the order of $10^{-4}$ per spin. For smaller $\lambda \lesssim 45$, the hopfion superstructure becomes unstable as the dense hopfions tend to merge with each other, as exemplified for $\lambda=38$ in Fig.~\ref{fig:meta}(b). In particular, for $\lambda \lesssim 41$ (red region), the spin state converges to meron quartet clusters, each of which consists of two merons with down-spin cores and two others with up-spin cores; two such meron clusters are found in the right most panel of Fig.~\ref{fig:meta}(b). Meanwhile, in the intermediate region of $41\lesssim \lambda \lesssim 45$ (yellow region), the optimization converges to a superstructure composed of $H = \pm 2$, instead of $H = \pm 1$, which is similar to the intermediate state shown in the middle panel of Fig.~\ref{fig:meta}(b). This suggests the possibility of hopfion superstructures with higher Hopf numbers. We note that skyrmions can also form superstructures with higher skyrmion numbers~\cite{Ozawa2017}. In contrast, for larger $\lambda \gtrsim 56$ (blue region), the hopfion superstructure becomes unstable, leaving a cluster state composed of four hopfions, as exemplified in the middle panel of Fig.~\ref{fig:meta}(c). In particular, for $\lambda \gtrsim 73$, such a cluster state becomes unstable and eventually changes into the meron cluster state similar to that we also observed in the red region in Fig.~\ref{fig:meta}(a), as shown in the right most panel of Fig.~\ref{fig:meta}(c) for $\lambda=76$.

The results suggest that once hopfions with $H=\pm 1$ are created with proper density and arrangemrent, one can expect self-organization of the 3D hopfion superstructures. Importantly, in a wide range of the initial density, the topological nature is preserved without relaxing to the ferromagnetic ground state, even though various superstructures can appear, as shown in Figs.~\ref{fig:meta}(b) and \ref{fig:meta}(c). Furthermore, it is interesting to note that higher-order hopfion superstructures can also be formed by fine-tuning of the hopfion density.

%%%%%%%%%%%%%%%%%%%%%%%%%%%%%%%%%%%%%%%%%%%%%%%%%%%%%%%%%%%%%%%%%%%%%%%%%%%%%%%%%%%%%%%%%%%%%%%%%%%%%%%%%%%%%%%%%%%%%%%%%%%%%%%%%%%%
\section{Summary and perspectives}
\label{sec:sum}
We have investigated the stability of magnetic hopfions in the 3D frustrated spin model. First, we showed that effective interactions between hopfions arranged horizontally can be understood by ``color rule": When hopfions are adjacent to each other so that their proximate spins are antiparallel (complementary colors), they feel repulsive forces and remain stable, whereas when they are adjacent with parallel spins (same colors), they cause fusion or pair annihilation. We also analyzed the interactions for vertically arranged hopfions, and found that a 1D periodic chain of hopfions can be stabilized by introducing the easy-plane anisotropy. Then, combining these findings, we successfully constructed the stable 3D hopfion superstructure with a staggered arrangement of the hopfion chains with $H = 1$ and $H = -1$ in a 2D square-lattice manner. Interestingly, any 2D cut perpendicular to the hopfion chains has a net skyrmion number with $N_{\rm sk} = 2$ per magnetic unit cell, which can be regarded as a magnetic meron lattice. This indicates that the superstructure is composed of the hopfion chains threaded by the meron strings. We also clarified that the hopfion superstructure is obtained as a metastable state in a particular range of the magnetic period. To the best of our knowledge, our finding provides the first example of at least metastable 3D crystalline superstructures of magnetic hopfions in spin systems. The detailed analyses suggest that the superstructure can be self-organized via the hopfion-hopfion interactions once hopfions are created with proper density and arrangement. We expect that it would be identified by experiments, such as neutron scattering and transport measurements, as higher harmonics peculiar to the hopfion superstructure and the topological Hall response from nonzero $N_{\rm sk}$.

Our results shed light on the possibility of stabilizing such a hopfion superstructure as the true ground state. A meron lattice similar to the 2D cut of our hopfion superstructure was found in the ground state of an effective spin model with long-range bilinear and biquadratic interactions derived from the coupling between itinerant electrons and localized spins~\cite{Hayami2022}. Given that such effective interactions are known to stabilize noncoplanar spin structures, such as skyrmions and hedgehogs (Bloch points)~\cite{Akagi2010,Akagi2012,Ozawa2016,Ozawa2017,Hayami2017,Okumura2020,Okumura2022}, the models with itinerant frustration~\cite{Hayami2021_itinfrustration} are worth studying. It would also be interesting to examine the effects of higher-order interactions, such as the sixth-order chiral-chiral interaction~\cite{Grytsiuk2020,Hayami2021_phaseshift} that is equivalent to the fourth-order differential term called the Skyrme term in the SF model~\cite{Faddeev1997}.

As a biproduct of our study on metastability, we discovered a superstructure with $H = \pm2$ when starting from the $H=\pm 1$ ones within a specific range of the magnetic period. This suggests an intriguing possibility of stabilizing superstructures with higher-order hopfions. Furthermore, it also implies the potential to convert a hopfion superstructure to the other one with a different Hopf number. These fascinating issues are left for future studies.

Topological spin textures have a hierarchical nature, presenting unique quantum phenomena and possible applications at each level. Our study revealed that magnetic hopfions, each of which consists of a twisted-ring like superstructure with magnetic skyrmions, can form a higher-level superstructure composed of the hopfion chanis. This finding would extend such a hierarchical structure in the topological spin textures and serve as a new platform for further exploration of topological quantum phenomena and dynamical properties.

%%%%%%%%%%%%%%%%%%%%%%%%%%%%%%%%%%%%%%%%%%%%%%%%%%%%%%%%%%%%%%%%%%%%%%%%%%%%%%%%%%%%%%%%%%%%%%%%%%%%%%%%%%%%%%%%%%%%%%%%%%%%%%%%%%%%

\section*{\label{acknowledge}Acknowledgments}
We thank R. Yambe for fruitful discussions. This work was supported by Grant-in-Aid for Scientific Research Grants (Nos. JP19H05822, JP19H05825, JP22K03509, JP22KJ0590, and JP24K22870) and JST CREST (No. JP-MJCR18T2). S. Kasai was supported by the Program for Leading Graduate Schools (MERIT-WINGS). The computation in this work has been done using the facilities of the Supercomputer Center, the Institute for Solid State Physics, the University of Tokyo.

%%%%%%%%%%%%%%%%%%%%%%%%%%%%%%%%%%%%%%%%%%%%%%%%%%%%%%%%%%%%%%%%%%%%%%%%%%%%%%%%%%%%%%%%%%%%%%%%%%%%%%%%%%%%%%%%%%%%%%%%%%%%%%%%%%%%

\bibliography{bibliography}

\end{document}